\documentclass{ws-mplb}
\usepackage{graphicx}
\newcommand{\e}{\mathbf}

\begin{document}

\markboth{N.~Voropajeva \& A.~Sherman}{Near-Boundary and Bulk Regions of a 2D Heisenberg Antiferromagnet}

%
\catchline{}{}{}{}{}
%

\title{NEAR-BOUNDARY AND BULK REGIONS OF A SEMI-INFINITE\\ TWO-DIMENSIONAL HEISENBERG ANTIFERROMAGNET}

\author{N.~VOROPAJEVA and A.~SHERMAN}

\address{Institute of Physics, University of Tartu, Riia 142, 51014 Tartu, Estonia}

\maketitle

\begin{history}
\received{(Day Month Year)}
\revised{(Day Month Year)}
\end{history}

\begin{abstract}
Using the spin-wave approximation elementary excitations of a semi-infinite two-dimensional $S=\frac12$ Heisenberg antiferromagnet are considered. The spectrum consists of bulk modes -- standing spin waves and a quasi-one-dimensional mode of boundary spin waves. These latter excitations eject bulk modes from two boundary rows of sites, thereby dividing the antiferromagnet into two regions with different dominant excitations. As a result absolute values of nearest-neighbor spin correlations on the edge exceed the bulk value.
\end{abstract}


\section{Introduction}
It is well known that at certain conditions defects of crystal
structure can generate states localized in the defect region, while
the magnitude of bulk states is suppressed in this region.\cite{lifshits,maradudin} This leads to the situation in which the
defect neighborhood and the rest of the crystal constitute two
systems with different dominant excitations. A crystal surface
can be also considered as a defect.\cite{prutton} As applied to the
surface, the mentioned situation leads to the appearance of a
near-boundary region whose properties differ from the bulk properties.

The influence of boundaries on the spectrum and observables of the quantum Heisenberg antiferromagnet has been studied in two\cite{hoglund,metlitski,pardini} (2D) and three\cite{voropajeva} (3D) dimensions. In particular it was shown that absolute values of the nearest-neighbor spin correlations near the boundary exceed the bulk
value. This result was interpreted as a manifestation of increased valence-bond-solid correlations near the edge.\cite{metlitski,pardini}

In this Letter, we propose another interpretation of the increased spin correlations near the edge. We relate their appearance to the peculiar spectrum of the semi-infinite antiferromagnet. The
spectrum involves bulk modes -- standing spin waves and a
quasi-one-dimensional mode of boundary spin waves. These latter excitations are observed in the two boundary rows of sites, and they eject the bulk excitations from this region. Thus the antiferromagnet appears to be divided into two regions with different dominant excitations. As known, nearest-neighbor spin correlations of the one-dimensional (1D) antiferromagnet are larger than in the 2D case. As a consequence the quasi-1D near-boundary mode produces larger spin correlations than the 2D bulk modes, which explains the observed\cite{hoglund,metlitski,pardini} increased near-boundary correlations. Similar interpretation can be applied to the 3D case.\cite{voropajeva} As will be seen below, the description of perturbations introduced by the edge into the magnon spectrum is in many respects similar to the problem of a local defect in a crystal.\cite{lifshits,maradudin} Thus seemingly different imperfections of crystal structure are described in the framework of the same approach.

\section{Model and its Elementary Excitations}
We suppose that the edge is located along one of the crystallographic axes and choose coordinates so that the antiferromagnet is situated in the half-space $l_x\ge0$ and described by the Hamiltonian
\begin{equation}\label{1}
    H=J\sum_{l_y}\sum_{l_x\geqslant0}\left(\e S_{l_y+1,l_x}\e S_{l_y l_x}+\e S_{l_y,l_x+1}\e S_{l_y l_x}\right),
\end{equation}
where sites of a 2D square lattice are labeled by the two coordinates $l_x$ and $l_y$, the lattice spacing is set as the unit of length, and $\e{S}_{l_y l_x}$ is the spin-$\frac12$ operator.

For the temperature $T=0$ the antiferromagnet has the long-range order and its elementary excitations can be described in the spin-wave approximation,
\begin{equation}\label{2}
    S^z_{\e{l}}=e^{i\e{\Pi
    l}}\left(\frac12-b^\dag_\e{l}b_{\e{l}}\right), \quad
    S^\pm_\e{l}=P^{\pm}_\e{l}b_\e{l}+P_\e{l}^{\mp}
    b^\dag_\e{l},
\end{equation}
where $\e{\Pi}=(\pi,\pi)$, $P^\pm_\e{l}=\frac12\left(1\pm e^{i\e{\Pi l}}\right)$, and $\e{l}=(l_x,l_y)$. The spin-wave operators $b_\e{l}$ and $b^\dag_\e{l}$ satisfy the Boson commutation relations. Substituting Eq.~\eqref{2} into Hamiltonian~\eqref{1}, dropping terms containing more than two spin-wave operators and constant terms, we obtain
\begin{eqnarray}\label{3}
H&=&J\sum_{k_y}\sum_{l_x\geqslant0}\biggl[2\left(1-\frac14\delta_{l_x 0}\right)b^\dag_{k_y l_x}b_{k_y l_x}
+\cos(k_y)\left(b_{k_y l_x}b_{-k_y,l_x}+b^\dag_{k_y l_x}b^\dag_{-k_y,l_x}\right)\nonumber\\
&&+\left(b_{k_y l_x}b_{-k_y,l_x+1}+b^\dag_{k_y l_x}b^\dag_{-k_y,l_x+1}\right)\biggr].
\end{eqnarray}
Here the translational invariance of the Hamiltonian along the $Y$ axis was taken into account and the Fourier transformation
$b_{k_y l_x}=N^{-1/2}\sum_{l_y}e^{i k_y l_y}b_{l_y l_x}$
was used with $N$ the number of sites in the $Y$ direction and $k_y$ the 1D wave vector.

To investigate the spectrum of elementary excitations we introduce
the two-component operator
$$
\hat B_{k_y l_x}=\left(
                         \begin{array}{c}
                           b_{k_y l_x} \\
                           b^\dag_{-k_y,l_x} \\
                         \end{array}
                   \right)
$$
and define the matrix retarded Green's function
\begin{equation}\label{4}
    \hat{D}(k_y t l_x l'_x)=
    -i\theta(t)\left\langle\left[\hat B_{k_y l_x}(t),\hat B^\dag_{k_y l'_x}\right]\right\rangle.
\end{equation}
In Eq.~\eqref{4}, $\hat B_{k_y l_x}(t)=e^{i H t}\hat B_{k_y l_x}e^{-i H t}$ with the Hamiltonian determined by Eq.~\eqref{3} and the angular brackets denote the statistical averaging.

To calculate Green's function~\eqref{4} we use the equation of motion,
\begin{eqnarray}\label{5}
i\frac{d}{dt}\hat D(k_y t l_x l'_x)&=&\delta(t)\delta_{l_x l'_x}\hat\tau_3
+J\biggl[2\left(1-\frac14\delta_{l_x 0}\right)\hat\tau_3
+\cos(k_y)\hat\tau_1\biggr]\hat D(k_y t l_x l'_x) \nonumber\\
&&+\frac J2\hat\tau_1\Bigl[\hat D(k_y t,l_x+1,l'_x)
+\hat D(k_y t,l_x-1,l'_x)\Bigr],
\end{eqnarray}
where
$$
    \hat\tau_1=\left(
                 \begin{array}{cc}
                   0 & 1 \\
                   -1 & 0 \\
                 \end{array}
               \right), \quad
     \hat\tau_3=\left(
                 \begin{array}{cc}
                   1 & 0 \\
                   0 & -1 \\
                 \end{array}
               \right).
$$
The solution of Eq.~\eqref{5} can be written in the form
\begin{equation}\label{6}
\hat D(k_y t l_x l'_x)=\hat D^{(0)}(k_y t l_x l'_x)
-\frac J2\int\limits_{-\infty}^\infty dt' \hat D^{(0)}(k_y,t-t',l_x 0)\hat D(k_y t' 0 l'_x),
\end{equation}
where $\hat {D}^{(0)}\left( {k_y tl_x {l}'_x } \right)$ is Green's function corresponding to Hamiltonian~\eqref{3} without the term proportional to $\delta _{l_x 0}$. After the Fourier transformation, $\hat{D}(k_y \omega l_xl'_x)=\int_{-\infty}^\infty e^{i\omega t}\hat{D}(k_ytl_xl'_x)\,dt$, and some mathematical manipulation we find
\begin{eqnarray}\label{7}
\hat D(k_y\omega l_x l'_x)&=&\hat D^{(0)}(k_y\omega l_x l'_x)\nonumber\\
&&-\frac J2 \hat D^{(0)}(k_y\omega l_x 0)
\left[\hat\tau_0+\frac J2 \hat D^{(0)}(k_y\omega 0 0)\right]^{-1}\hat D^{(0)}(k_y\omega 0 l'_x).
\end{eqnarray}
Here $\hat{\tau}_0$ is a $2\times 2$ identity matrix. It is noteworthy that, except for the matrix form and the parametric dependence on $k_y$, Eq.~\eqref{7} is similar in form to the equation for Green's function of a crystal with a local defect.\cite{lifshits,maradudin} As in this latter equation, the first term in the right-hand side of Eq.~\eqref{7} describes bulk excitations, while the poles of the second term correspond to excitations localized in the defect region.

To calculate Green's function $\hat D^{(0)}(k_y\omega l_x l'_x)$ it is necessary to diagonalize Hamiltonian~\eqref{3} without the term proportional to $\delta_{l_x 0}$. This can be fulfilled using the Bo\-go\-liubov-Tyablikov transformation,\cite{tyablikov}
\begin{equation}\label{8}
    \beta_{k_y k_x}=\sum_{l_x\geqslant0}(u^{*}_{k_y l_x k_x} b_{k_y l_x}-v_{-k_y,l_x k_x}b^\dag_{-k_y, l_x}),
\end{equation}
where the coefficients $u_{k_y l_x k_x}$ and $v_{k_y l_x k_x}$ satisfy
the usual orthonormality conditions which follows from the Boson commutation relations of the operators $b_{k_y l_x}$ and $\beta_{k_y k_x}$. From these conditions and from the requirement that the Hamiltonian be diagonal in the representation of operators $\beta_{k_y k_x}$ we obtain the following system of equations for the coefficients $u_{k_y l_x k_x}$, $v_{k_y l_x k_x}$ and the energy of elementary excitations $E_{k_y k_x}$:
\begin{eqnarray}
E_{k_y k_x}u^{*}_{k_y l_x k_x}&=&
2Ju^{*}_{k_y l_x k_x}+J\cos(k_y)v_{-k_y,l_x k_x}\nonumber\\
&&+\frac J2\bigl(v_{-k_y,l_x+1,k_x}+v_{-k_y,l_x-1,k_x}\bigr), \nonumber\\[-0.5ex]
&&\label{9}\\[-0.5ex]
-E_{k_y k_x}v_{-k_y,l_x k_x}&=&
2Jv_{-k_y,l_x k_x}+J\cos(k_y) u^{*}_{k_y l_x k_x}\nonumber\\
&&+\frac J2\bigl(u^{*}_{k_y,l_x+1,k_x}+u^{*}_{k_y,l_x-1,k_x} \bigr),\nonumber
\end{eqnarray}
with the boundary conditions
\begin{equation}\label{10}
    u^{*}_{k_y,l_x=-1,k_x}=0, \quad v_{-k_y,l_x=-1,k_x}=0.
\end{equation}
Solutions of Eqs.~\eqref{9} and \eqref{10} are standing waves,
\begin{eqnarray}
u^{*}_{k_y l_x k_x}&=&A_{k_y k_x}\sin[k_x(l_x+1)],\nonumber\\
v_{-k_y,l_x k_x}&=&B_{k_y k_x}\sin[k_x(l_x+1)],\nonumber\\
A_{k_y k_x}&=&\sqrt{\frac{2}{\pi}}\frac{2J+E_{k_y k_x}}{\sqrt{\left(2J+E_{k_y k_x}\right)^2-\left(2J\gamma_{k_y k_x}\right)^2}},\label{11}\\
B_{k_y k_x}&=&-\sqrt{\frac{2}{\pi}}\frac{2J\gamma_{k_y k_x}}{\sqrt{\left(2J+E_{k_y k_x}\right)^2-\left(2J\gamma_{k_y k_x}\right)^2}},\nonumber\\
E_{k_y k_x}&=&2J\sqrt{1-\left(\gamma_{k_y k_x}\right)^2},\nonumber
\end{eqnarray}
where $\gamma_{k_y k_x}=\frac12[\cos(k_x)+\cos(k_y)]$ and $k_x$ varies continuously in the range $(0,\pi)$. Using solutions \eqref{11} we find after the Fourier transformation
\begin{eqnarray}
&&\hat D^{(0)}(k_y\omega l_x l'_x)=
\!\!\int_0^\pi \!\! \sin[k_x(l_x+1)\sin[k_x(l'_x+1)]\nonumber\\
&&\quad\times\left(\frac{\hat P_{k_y k_x}}{\omega-E_{k_y k_x}+i\eta}-\frac{\hat Q_{k_y k_x}}{\omega+E_{k_y k_x}+i\eta}\right)dk_x,\nonumber\\[-0.5ex]
&&\label{12}\\[-0.5ex]
&&\hat P_{k_y k_x}=\left(
                   \begin{array}{cc}
                     A^2_{k_y k_x} & A_{k_y k_x}B_{k_y k_x} \\
                     A_{k_y k_x}B_{k_y k_x} & B^2_{k_y k_x} \\
                   \end{array}
                 \right),\nonumber\\
&&\hat Q_{k_y k_x}=\left(
                   \begin{array}{cc}
                     B^2_{k_y k_x} & A_{k_y k_x}B_{k_y k_x} \\
                     A_{k_y k_x}B_{k_y k_x} & A^2_{k_y k_x} \\
                   \end{array}
                 \right),\nonumber
\end{eqnarray}
where $\eta=+0$.
\begin{figure}
\begin{center}
\includegraphics[width=0.45\linewidth]
{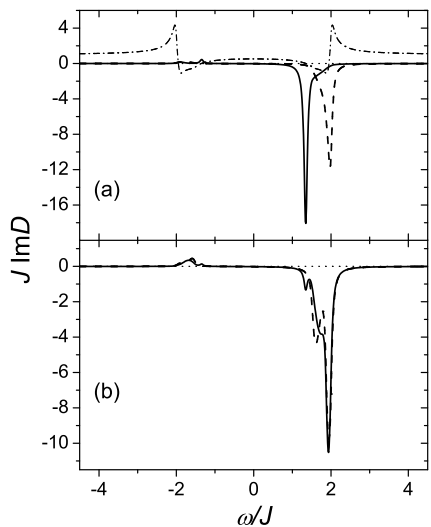} \caption{The imaginary parts of Green's functions
$D_{11}(k_y\omega l_x l_x)$ (the solid lines) and $D_{11}^{(0)}
(k_y\omega l_x l_x)$ (the dashed lines) for $l_x =0$ (a) and $l_x
=2$ (b). $k_y=0.6\pi$. In part (a), the dash-dotted line demonstrates the real part of the denominator in the second term in the right-hand side of Eq.~\protect\eqref{7}.}\label{fig1}
\end{center}
\end{figure}
\begin{figure}
\begin{center}
\includegraphics[width=0.45\linewidth]
{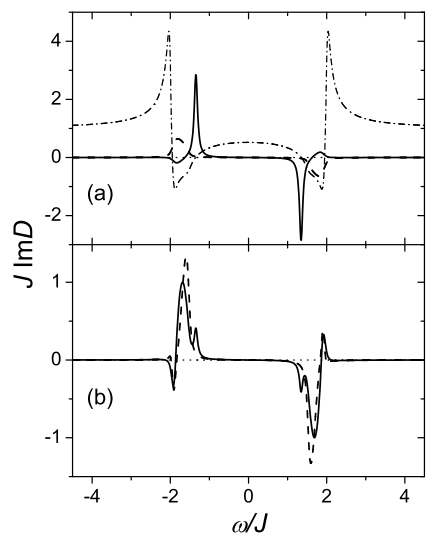} \caption{The same as in Fig.~\ref{fig1} for Green's
functions $D_{12}(k_y\omega l_x l_x)$ and $D_{12}^{(0)} (k_y\omega
l_x l_x)$.}\label{fig2}
\end{center}
\end{figure}

The poles of Green's function~\eqref{12} correspond to bulk excitations -- standing spin waves~\eqref{11}. Apart from them Green's function $\hat D(k_y\omega l_x l'_x)$ may have poles connected with the second term in the right-hand side of Eq.~\eqref{7}. The imaginary parts of Green's functions $\hat D^{(0)}(k_y\omega l_x l'_x)$ and $\hat D(k_y\omega l_x l'_x)$ are shown in Figs.~\ref{fig1} and~\ref{fig2} for different distances from the edge. On the edge, $l_x=0$, the spectrum $\mbox{Im}\hat D(k_y\omega l_x l'_x)$ is dominated by the peak arising from this second term. Indeed, as seen from Figs.~\ref{fig1}(a) and~\ref{fig2}(a), the peak frequency coincides with a zero of the denominator in the term. The standing spin waves manifest themselves as a weak shoulder on the high-frequency side of the peak in Fig.~\ref{fig1}(a) -- the excitations connected with the peak eject bulk modes from the edge row of sites. The similar situation is observed in the second row. However, as seen in Figs.~\ref{fig1}(b) and~\ref{fig2}(b), already in the third row the peak is weak and the spectrum is dominated by the continuum of standing waves. Thus, the antiferromagnet is divided into two regions with different spin excitations. The two rows near the edge are the location of the mode connected with the pole of the second term in Eq.~\eqref{7}. This mode has the dispersion $\omega(k_y)=\sqrt2 J\left|\sin\left(k_y\right)\right|$ and is termed the boundary spin-wave mode.\cite{disp} Near $k_y=0$ and $k_y=\pi$ it is ill-defined -- the real part of the denominator of the second term is small but nonzero. Excitations of the rest of the crystal are the standing spin waves with dispersion \eqref{11}. This picture is in many respects similar to the situation in the problem of a local defect:\cite{lifshits,maradudin} if local states arise near the defect, they eject bulk states from the defect region.

\section{Spin Correlations}
Nearest-neighbor spin correlations can be expressed in terms of correlations of spin-wave operators using Eq.~\eqref{2} and the translation invariance of Hamiltonian~\eqref{1} along the $Y$ axis,
\begin{eqnarray}\label{13}
\langle \e S_{\e l}\e S_{\e l'}\rangle&=&
\frac{1}{2N}\sum_{k_y}\Bigl\{2\cos[k_y(l_y-l'_y)]\langle b_{k_y l_x}b_{-k_y,l'_x}\rangle\nonumber\\
&&+\langle b^\dag_{k_y,l_x}b_{k_y,l_x}\rangle+\langle b^\dag_{k_y,l'_x}b_{k_y,l'_x}\rangle\Bigr\}-\frac14.
\end{eqnarray}
Bearing in mind the property of Green's function~\eqref{7} $D_{ij}(k_y\omega l_x l'_x)=D_{ji}(k_y\omega l'_x l_x)$,
the spin-wave correlations in Eq.~\eqref{13} can be expressed as
\begin{equation}\label{14}
    \left\langle \hat B_{k_y l_x}\hat B^\dag_{k_y l'_x}\right\rangle=
    \int_{-\infty}^\infty \frac{d\omega}{\pi}\frac{\mathrm{Im}[\hat D(k_y \omega l_x l'_x)]}{e^{-\omega/T}-1}.
\end{equation}
\begin{figure}
\begin{center}
\includegraphics[width=0.6\linewidth]
{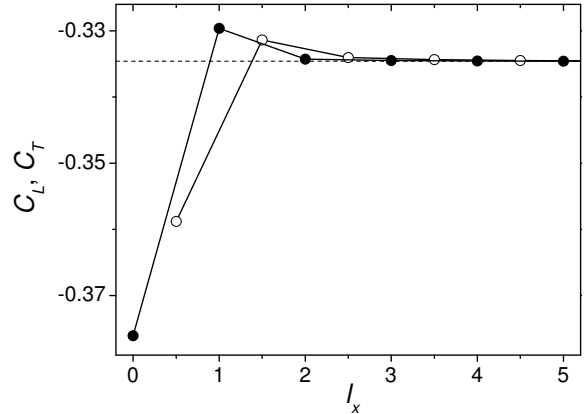} \caption{The nearest-neighbor spin correlations
parallel (filled circles) and perpendicular (open circles) to the
edge as functions of the distance $l_x$ to it for $T=0$. Bonds perpendicular to the edge have their centers at half-integer values of $l_x$. Solid lines are the guide to the eye. The dashed line indicates the bulk value.}\label{fig3}
\end{center}
\end{figure}

The calculated nearest-neighbor spin correlations parallel and
perpendicular to the edge, $C_L=\left\langle \e S_{l_y+1,l_x}\e
S_{l_y l_x}\right\rangle$ and $C_T=\left\langle \e S_{l_y,l_x+1}\e
S_{l_y l_x}\right\rangle$, are shown in Fig.~\ref{fig3}. As seen from
the figure, main deviations from the bulk value of the correlations
fall on the edge and the second to the edge row of spins, i.e.\ on
the existence domain of the boundary mode. The largest in absolute value spin correlations are observed on the edge and between the edge and the second to the edge row. The main contribution to these large
correlations is made by the boundary mode which has quasi-1D
character. This result conforms with the fact that the modulus of
nearest-neighbor spin correlations in the 1D antiferromagnet\cite{izyumov}
($|\!\left\langle\e{S}_l\e S_{l+1}\right\rangle\!|=0.4432$) exceeds its value in the 2D case
($|\!\left\langle\e{S_l} \e{S_{l+a}}\right\rangle\!|=0.3346$ in our
calculations, which is close to the values obtained earlier;\cite{hamer,gochev} $\e a$ is the 2D vector connecting neighbor
sites). Thus, we relate the enhanced spin correlations on the edge
to the separation of the crystal into two regions with essentially
different excitations -- the boundary region with the quasi-1D mode
and the bulk region with the 2D spin waves. The correlations in
the second and between the second and third rows are smaller than in
the bulk due to the destructive summation of the contributions of
these two types of excitations [see Eq.~\eqref{7}]. Qualitatively
the obtained picture of spin correlations is similar to that
observed in the 3D case.\cite{voropajeva} However, in the 2D case
the deviation of the boundary correlations from the bulk value is
larger than for 3D -- 12\% in comparison with 5\% in the latter
case.

If results in Fig.~3 are compared with the data of Monte Carlo simulations\cite{hoglund} two differences stand out: i) in our calculations, the absolute value of the edge correlation on the parallel bond is larger than on the bond perpendicular to the edge, while in the Monte Carlo data the relation is opposite; ii) there are weak oscillations of spin correlations around the bulk value which are perceptible over a few lattice periods from the edge in the Monte Carlo data, while such oscillations are missing from our results. Partly, these differences can be ascribed to the dissimilarity of the used samples -- a square-shaped finite crystal in the Monte Carlo simulations and a semi-infinite crystal in our case. However, we suppose that the main reason for these differences is the quasi-1D character of the boundary mode. This mode is the origin of the mentioned peculiarities of spin correlations. However, as known, the spin-wave approximation is unsuitable for the 1D antiferromagnet. Therefore, it is believed that the approach used gives only a rough description of the boundary excitations. It would be interesting to compare our results\cite{voropajeva} for the 3D case, when both the boundary and bulk modes are satisfactorily described by the spin-wave approximation, with Monte Carlo data. However, to our knowledge such simulations are lacking.

\section{Conclusion}
In this Letter, we relate the increased spin correlations near
the edge of the semi-infinite spin-$\frac 12$ 2D antiferromagnet to its peculiar spectrum of spin excitations. The spectrum involves bulk
modes -- standing spin waves and the quasi-1D boundary mode. This mode
is observed in the two boundary rows of sites, and it ejects the
bulk modes from this region. Thereby the antiferromagnet is divided
into two regions with different dominant excitations and correlations.
The mathematical description of the boundary mode and its behavior
has much in common with the problem of a local defect in a crystal.

\section*{Acknowledgments}
This work was supported by the ETF grant No. 6918.


\begin{thebibliography}{00}
\bibitem{disp} For the surface mode of the 3D semi-infinite antiferromagnet on a square lattice we found\cite{voropajeva} the dispersion $\omega(k_y k_z)=\sqrt{\mathstrut d(d-1)}J\sqrt{1-(\gamma_{k_y k_z})^2}$ for the surface perpendicular to the $X$ axis. Here the dimensionality $d=3$. The above dispersion of the boundary mode in the case $d=2$ also satisfies this formula.
\end{thebibliography}
\end{document}